# Non-equilibrium character of resistive switching and negative differential resistance in Ga-doped $Cr_2O_3$ system

R.N. Bhowmik[*], and K.Venkata Siva

Department of Physics, Pondicherry University, R. V. Nagar, Kalapet, Pondicherry-605014, India.

[*]Corresponding author: Tel.: +91-9944064547

E-mail: rnbhowmik.phy@pondiuni.edu.in

**Abstract:** The samples of Ga-doped $Cr_2O_3$ system in rhombohedral crystal structure with space group $R\bar{3}C$ were prepared by chemical co-precipitation route and annealing at 800 $^0$C. The current-voltage (I-V) curves exhibited many unique non-linear properties, e.g., hysteresis loop, resistive switching, and negative differential resistance (NDR). In this work, we report non-equilibrium properties of resistive switching and NDR phenomena. The non-equilibrium I-V characteristics were confirmed by repetiting measurement and time relaxation of current. The charge conduction process was understood by analysing the I-V curves using electrode-limited and bulk-limited charge conduction mechanisms, which were proposed for metal electrode/metal oxide/metal electrode structure. The I-V curves in the NDR regime and at higher bias voltage in our samples do not obeyed Fowler-Nordheim equation, which was proposed for charge tunneling mechanism in many thin film junctions. The non-equilibrium I-V phenomena were explained by considering the competitions between the injection of charge carriers from metal electrode to metal oxide, the charge flow through bulk material mediated by trapping/de-trapping and recombination of charge carriers at the defect sites of ions, the space charge effects at the junctions of electrodes and metal oxides, and finally, the out flow of electrons from metal oxide to metal electrode.

# 1. Introduction:

Recently, different types of random access memory (RAM) devices, e.g., ferroelectric random access memory (FeRAM), magnetoresistive random access memory (MRAM), and phase change random access memory (PRAM), have been introduced in spintronics [1]. There is an attempt to develop non-volatile memory (NVM) devices where data can be stored /written by magnetic process and stored information can be retrieved/read by electrical process. MRAM is one of the novel approaches that can store data in magnetic storage element, called magnetic tunnel junction (MTJ). MRAM has almost all the criteria of becoming the next generation memory devices (high-speed reading and writing capacity with additional non-volatility). These NVM devices utilize charge-spin coupling in the material and such materials are not easily available in nature. In an attempt to search alternative materials for non-volatile memory device applications, insulator-magnetic metal oxides with high resistive switching have drawn a world-wide attention [1-2]. This is due to the fact that resistive Random Access Memory (ReRAM) devices are based on resistive switching (RS) process; resistive state is controlled by sweeping electric voltage or current pulses and retains data even at power shut down condition [3-5]. ReRAM is believed to be a potential candidate for the applications in next-generation NVM devices due to its low power consumption, high switching speed and high density [6-7].

The thin films of transition metal oxides, e.g., $Cr_2O_3$ [7], $Ga_2O_3$ [8-9], $Al_2O_3$ [10], $Fe_2O_3$ [11], Ga-doped $Fe_2O_3$ [12], $Ba_{0.7}Sr_{0.3}TiO_3$ [13], NiO [14], $TiO_2$ [15-16], and $SiO_2$ [17] have shown different types of resistive switching, i.e., unipolar, bipolar or abnormal nature. The unipolar switching is polarity-independent (symmetric) and resistance state of the device is switched (LRS ↔ HRS) by consecutive application of electric field of either in same or opposite directions. In bipolar switching, resistance state of the device is polarity-dependent (antisymmetric) and it switches LRS→HRS→ HRS →LRS by applying the successive electric fields with alternate polarity, i.e., ON state (LRS) at one voltage polarity and OFF state (HRS) on reversing the voltage polarity. In abnormal bipolar resistive switching, the resistance state of the device is switched from LRS to HRS during positive polarity of electric field, and LRS to HRS during negative polarity of electric field. The abnormal resistive switching phenomenon has been observed in devices like Pt/GaO$_3$/Pt [8] and Pt/TiO$_2$/Pt [18]. Recently, thin films of BaTiO$_3$ [19] and ZnO [20] have shown the coexistence of resistive switching and negative differential resistance (NDR) behaviour ($\frac{dV}{dI} < 0$). The materials with room temperature NDR and resistive swirching are of increasing interest for applications in analog and digital circuits, including logic gates, voltage-controlled oscillator, and flip-flop circuit [21]. The additional feature of I-V loop is also important for applications of the materials in switching, low power memory, dynamic random access memory, static random access memory, and high storage density with long retention [22].

We present resistance switching and NDR phenomena at room temperature in a new material, developed by incorporating non-magnetic Ga atoms in the antiferromagnetic α-Cr$_2$O$_3$. The magnetic measurement [23] has shown Ga doped α-Cr$_2$O$_3$ system as a diluted antiferromagnet with ordering temperature at about 50 K and a typical paramagnet at room temperature. This work is devoted to study the non-equilibrium I-V properties in thin pellet shaped samples of polycrystalline Ga doped α-Cr$_2$O$_3$ system using Pt/CrGaO/Pt structure. The present transition metal oxide samples are electrically high resistive semiconductor or insulator; where electric field induced migration and concentration gradient induced diffusion of oxygen ions/vacancies can play a crucial role on electronic properties. Hence, the I-V study for such material under high electric voltage will be useful to get insight of the role of microscopic level lattice defects on charge conduction mechanism. The non-equilibrium properties of I-V characteristics will be understood using experimental results of bias voltage cycling and time relaxation effects on current measurements.

## 2. Experimental

### 2.1. Material preparation and characterization

Details of the preparation of Ga doped $Cr_2O_3$ samples by chemical co-precipitation and structural phase stabilization have been described in earlier work [23]. In the present work, we used the samples with chemical compositions $Cr_{1.45}Ga_{0.55}O_3$ and $Cr_{1.17}Ga_{0.83}O_3$. The X-ray diffraction pattern confirmed single phased rhombohedral structure with profile fitted cell parameters ($a$ = 4.955 Å, $c$ = 13.572 Å, V = 288.5 Å$^3$ for $Cr_{1.45}Ga_{0.55}O_3$, $a$ = 4.955 Å, $c$ = 13.573 Å, V = 288.7 Å$^3$ for $Cr_{1.17}Ga_{0.83}O_3$). The grain size of the samples (~50 nm and 44 nm for $Cr_{1.45}Ga_{0.55}O_3$ and $Cr_{1.17}Ga_{0.83}O_3$, respectively) was found in same range. Raman spectra of the Ga doped samples were matched to the structure of α-$Cr_2O_3$. This provided microscopic evidence of the occupancy of Ga atoms into the sites of Cr atoms in α-$Cr_2O_3$.

### 2.2. Measurement of I-V characteristics

Current-Voltage (I-V) characteristics of the samples were measured using Keithley dual Source Meter (2410-C, USA) and high resistance meter (6517B, Keithley, USA). The pellet shaped samples (∅~13 mm, t~0.3-0.5 mm) were sandwiched between two Pt electrodes using pressure in home-made sample holder to make a vertical Pt/CrGaO/Pt device structure. Two modes ($P$ and $N$) with 4 segments were adopted to measure the I-V curves by sweeping bias voltage from 0 to suitable voltage limit across the sample. In P mode, voltage was swept in positive side with step sequences $\overline{0 \to +V_{max}}(S1) \to \overline{+V_{max} \to 0}(S2) \to \overline{0 \to -V_{max}}(S3) \to \overline{-V_{max} \to 0}(S4)$. In N mode, bias voltage was swept in negative side with sequences $\overline{0 \to -V_{max}}(S1) \to \overline{-V_{max} \to 0}(S2) \to \overline{0 \to +V_{max}}(S3) \to \overline{+V_{max} \to 0}(S4)$. Reproducibility of the I-V features and charge relaxation effect was tested by additional protocols, (1) 20 times repetition (20P or 20N) of loop measurement; (2) sweeping the voltage with sequences $+V_{max} \to 0 \to -V_{max}$ and $-V_{max} \to 0 \to +V_{max}$; (3) varying the delay time between two set voltage points; and (4) time dependent current (I) measurement at constant voltage.

## 2.3. Method of I-V curve analysis

The I-V characteristics in most of the metal oxides using M/MO/M structure depends on two mechanisms [24-25]. First one is the electrode-limited conduction (ELC) mechanism, which includes Schottky emission and Fowler-Nordheim tunneling. Second one is the bulk-limited conduction (BLC) mechanism, which includes space charge limited conduction (SCLC) and Pool-Frenkel (P-F) emission. The ELC mechanism depends on the electrical properties at electrode -material interfaces, whereas BLC mechanism relies upon electrical properties of the material itself. The Schottky emission mechanism is described by following equation.

$$J = A * T^2 \exp\left[\frac{-q(\varphi_B - \sqrt{\frac{qE}{4\pi\varepsilon}})}{k_B T}\right] \quad (1)$$

J is the current density, $t$ is the sample thickness, V is applied voltage, A is the Richardson constant, E (= V/t) is the electric field, T is measurement temperature (300 K), $\varphi_B$ is barrier height, $\varepsilon$ (= $\varepsilon_0 \varepsilon_r$) is the dynamic dielectric constant, and $k_B$ is Boltzmann constant. The validity of Schottky emission can be tested from a linear fit in the lnJ vs. $\sqrt{E}$ (or lnI vs. $\sqrt{V}$) plot. If $\varphi_B$ at the M/MO junction is much higher than kinetic energy of incoming electrons, tunnelling of electrons through barrier is controlled by Fowler-Nordheim equation [26].

$$J = AE^2 e^{(-\frac{B}{E})} \quad \text{where } A = \frac{q^3}{8\pi h \varphi_B}, B = \frac{8\pi\sqrt{2m^*q}(\varphi_B)^{\frac{3}{2}}}{3h}; \quad E = V/t \quad (2)$$

In case of SCLC mechanism, the Power law: $I \sim V^m$ with exponent $m = 2$ is applicable for the material without any trapping effect between the electrodes [27]. The trapping and de-trapping of the electrons at defect sites, while hopping/transporting through metal oxide, are controlled by P-F emission that satisfies the following equation.

$$J = q n_c \mu \left(\frac{V}{t}\right) \exp\left[\frac{-q\varphi_B}{k_B T} + \frac{q\sqrt{\frac{qV}{\pi\varepsilon t}}}{r k_B T}\right] \quad (3)$$

$n_C$ is the density of charge carriers in the conduction band, $\mu$ is the electronic drift mobility and $r$ is the coefficient (range: 1-2). The validity of P-F emission gives a linear fit in the plot of ln(J/V) vs. $\sqrt{V}$. In P-F emission mechanism, the positively charged oxygen vacancies act as trapping centers. The electrons hopping between oxygen vacancies correspond to HRS in I-V curve. At higher voltage, energy barrier for the transport of electrons is reduced and the oxygen vacancies migrate towards the negative electrode (cathode). The electric field induced de-trapping of the electros from traps into the conduction band increases $n_C$ and current.

## 3. Experimental Results

### 3.1 Analysis of basic I-V characteristics

We have measured the I-V curves within three bias voltage limits ± 50 V, ±100 V and ± 200 V and analyzed the data using existing models. However, plots and data analysis of the samples are not shown for all the voltage ranges to optimise the number of figures. The I-V characteristics and analysis of α-$Cr_{1.45}Ga_{0.55}O_3$ sample have been depicted for voltage limits ± 50 V and ± 200 V. Fig. 1 ($V_{max}$= ± 50 V) has shown I-V data for P (a), N (b) and 20P (c) modes. On following the measurement sequences (S1→S2→S3→S4) of P mode, I-V curve initially showed a rapid increase up to a peak value ($I_P$) at bias voltage $V_P$ at low voltage regime of segment S1, followed by the appearance of NDR regime on further increase of the voltage up to +50 V. In segment S2, the current gradually decreases on decreasing the bias voltage from +50 V to 0 V. On reversing the direction of bias voltage, I-V characteristics in segments S3 and S4 are identical to that in segments S1 and S2, except reversal of the directions of I and V. The I-V curves in N mode showed similar character as in P mode, but the loop area in negative bias side is smaller than that in positive bias side. This means the loop area formed by reversing the polarity of bias voltage is asymmetric for N mode, unlike a symmetric loop in P mode for both positive and negative bias voltages. The symmetric loop area in case of P mode is reproduced after 20 times repetition of the measurement, although current slowly decreased. The inset of Fig. 1(c) showed a rapid decrease of loop area for initial 2-3 repetitions and later on less sensitive on repeating the I-V loop measurement. The R-V loops, where R = V/I, also confirms the symmetric and asymmetric shapes in the P (Fig. 1(d) and N (Fig. 1(e)) modes, respectively. The R-V curves suggested abnormal bipolar switching of the resistance state from LRS to HRS on reversing the bias voltage direction for both positive and negative voltage [8, 18, 28-29]. A gap in the R-V curves near to zero bias voltage between S1 and S2 or S3 and S4 segments is associated to a sudden jump of the resistance state LRS→HRS and HRS→LRS due to reversing the bias voltage path on same side or opposite side. This shows non-volatile (irreversible) resistive states in the material. Fig. 1(f) shows the voltage dependence of differential conductance (dI/dV) to confirm NDR effect. The dI/dV vs. V curves show that I-V curve is not exactly linear (Ohmic behaviour) at the low voltage regime of S1 and S3 segments. The dI/dV curve achieved a peak before reaching to zero value at $I_P$ for bias voltage $V_P$. At V> $V_p$, dI/dV becomes negative and attained a minimum and subsequently return back towards zero or positive value on further increase of voltage in S1 segment. In S2 segment, dI/dV remained positive or near to zero till the bias voltage approaches towards zero value. Then dI/dV features during segments S3 and S4 are similar as in S1 and S2, and almost symmetric about the current axis except a minor difference in the magnitude.

We used positive bias region (segments 1 and 2 in P mode) of the I-V curve to analyse the charge conduction mechanism. Fig. 1(g) shows that I-V curves in the low voltage regime of the S1 segment and entire S2 segment followed power law ($I{\sim}V^m$) with exponent *m* in the range 1.05-2.37 and 0.92-1.13, respectively. For trapping free SCLC mechanism in solid state devices, the exponent is 2; otherwise, exponent values can be less than 2 [11-12, 30]. This means the I-V curve in the S2 segment does not follow a typical SCLC mechanism. To realize an appropriate charge conduction mechanism, we analyzed S2 segment of the P mode (Fig. 1(h)), since NDR effect was noted in the segment S1. The I-V curve (segment S2 of P mode) followed Schottky equation with slope 0.007 and P-F equation with slope 0.089 in the low voltage (10 V-40 V) and high voltage (40 V-50 V) regimes, respectively.

Fig. 2 plots the I-V curves and related parameters of α-$Cr_{1.45}Ga_{0.55}O_3$ sample for bias voltage limit ± 200 V and measured using P mode. Fig. 2(a) shows a nearly symmetric I-V loop and NDR effect for positive and negative sides of bias voltage. The R-V loop (Fig. 2(b)) confirmed the switching of resistive states from LRS to HRS on reversing the bias voltage direction for both positive (S1→S2) and negative polarity (S3→S4). A gap was observed between S1 and S2 or S3 and S4 segments as the bias voltage returned back to zero. The dI/dV vs. V curve in Fig. 2(c) indicated NDR effect with features similar to that observed for bias voltage within ± 50 V. The repetition of measurement within bias voltage ± 200 V has reproduced the I-V characteristics (Fig. 2(d)). The loop area, after an initial decrease, is well stabilized on further repeating the measurement. The associated resistance switching (Fig. 2(e)) is affected by relaxation effect of measurement current. The R(V) curve showed a sharp decrease to achieve a sharp minimum at $V_p$ and then rapidly increased in the NDR regime of the S1 segment of I-V curve. The resistance slowly increased on returning the bias voltage to zero in the S2 segment. The behaviour of the R(V) in the S3 and S4 segments of the I-V curve is identical to that in S1 and S2 segments. A gap in the R(V) near to zero bias voltage is observed on reversing the bias voltage. Fig. 2(f) shows that power law with exponent ~ 1.03 is applicable only in low voltage (up to +16 V) regime of the S1 segment. In the S2 segment, power law is applicable with variable exponents ~0.92 and ~ 1.13 for low voltage and high voltage regimes, respectively. As shown in Fig. 2(g), the I-V curve in segment S2 of P mode obeys Schottky equation with slope 0.004 and P-F equation with slope 0.009 in the low voltage (< 100 V) and high voltage (170 V-200 V) regimes, respectively.

I-V characteristics for α-$Cr_{1.17}Ga_{0.83}O_3$ sample within ± 50 V are remarkably different from that in α-$Cr_{1.45}Ga_{0.55}O_3$ sample. Fig. 3 shows that I-V loop in positive volatge is highly asymmetric (large loop area) in comparison to a small loop area in negative voltage. The loop area in negative voltage side is significantly reduced for N mode (Fig. 3(b)) in comparison to P mode (Fig. 3(a)). The I-V curves of α-$Cr_{1.17}Ga_{0.83}O_3$ sample do not show clear signature of NDR effect for bias voltage within ± 50 V. The current rapidly increased with voltage at initial stage (up to 15-20 V), followed by a weak increasing/minor decreasing trend at higher voltages in the S1 segment of P mode and S3 segment of N mode. The decreasing trend of current at higher voltages of S1 and S3 segments increases on repeating the measurement for 20 times (Fig. 3(c)). In S2 and S4 segments, the current slowly decreased towards zero on decreasing the bias voltage from ± 50 V. The loop area gradually decreased on increasing the repetition of measurement within voltage ± 50 V (inset of Fig. 3(c)). Similar decrease of loop area was observed for repeating the measurement within ± 200 V (not shown in figure). Interestingly, NDR appeared for α-$Cr_{1.17}Ga_{0.83}O_3$ sample on repeating the I-V measurement. R-V plot (Fig. 3(d-f)) showed a rapid decrease of resistance at lower voltages (LRS) and an increase of resistance at higher voltages in S1 and S3 segments. The sample switched its resistive state to HRS in S2 and S4 segments on reversing the direction of bias voltage. The resistance gap near to zero bias voltage is nearly zero for the first loop within ± 50 V and a gap appeared on repeating the measurement, suggesting a switching of the α-$Cr_{1.17}Ga_{0.83}O_3$ sample to higher resistance state. I-V curves in P mode (Fig. 3(g)) obeyed the power law with exponent (m) ~ 1.85 and ~ 0.16 for the low voltage (< 15 V) and high voltage (> 20 V) regimes, respectively, of the S1 segment. In the S2 segment, the exponents are ~1.70 (low voltage regime) and ~ 1.13 (high voltage regime). The Schottky equation with slope 0.010 and P-F equation with slope 0.055 are applicable at the low voltage and high voltage regimes, respectively for the I-V curve in segment S2 of P mode (Fig. 3(h)).

Fig. 4 plots the I-V characteristics of the α-$Cr_{1.17}Ga_{0.83}O_3$ sample within ± 100 V (Fig. 4 (a-b) and ± 200 V (Fig. 4(c)). We observe that overall loop area, as well as loop area in the negative voltage side, increased by increasing the bias voltage limit in comparison to the I-V curve measured within ± 50 V. NDR effect is now appeared for negative voltage side also. The NDR effect with symmetric loop is more prominent during voltage cycling of I-V curves within bias voltage ± 100 V and ± 200 V. This means I-V loop in α-$Cr_{1.17}Ga_{0.83}O_3$ sample is highly non-equilibrium in character. The prominence of NDR effect depends on the voltage limit and repetition of I-V loop measurements. The resistive switching from LRS (in S1 and S3 segments) to HRS in S2 and S4 segments is reproduced on reversing the direction of bias voltage during cycling. A small gap between LRS and HRS near to zero bias voltage is noted in the R-V loop for bias voltage limits ± 100 V and ± 200 V. Power law fit of the I-V curves within voltage limit ± 100 V (Fig. 4(d)) showed the exponent (m) ~ 1.36 in the low voltage (< 10 V) regime and ~ 0.29 in the high voltage (> 30 V) regime of S1 segment. In S2 segment, the exponents are ~1.00 (low voltage regime) and ~ 0.95 (high voltage regime). The I-V curve in segment S2 (Fig. 4(e)) is well fitted by Schottky equation with slope 0.006 and P-F equation with slope 0.020 in the low and high voltage regimes, respectively.

The power law exponent (m) in the low voltage range of S1 segment is found in the range 1.03-1.53 and 1.36-1.85 for the samples $Cr_{1.45}Ga_{0.55}O_3$ and $Cr_{1.17}Ga_{0.83}O_3$, respectively without any specific increasing or decreasing trend of variation with the increase of voltage limit from 50 V to 200 V. However, exponent value in the low voltage range of S2 segment has decreased from 1.06 to 0.92 for $Cr_{1.45}Ga_{0.55}O_3$ sample and 1.70 to 0.97 for $Cr_{1.17}Ga_{0.83}O_3$ sample for increasing voltage limit 50 V to 200 V. The *m* at higher voltages of S2 segment is found in the range of 0.92-1.13 (increasing trend) for $Cr_{1.45}Ga_{0.55}O_3$ sample and 0.95-1.13 (without any specific trend) for $Cr_{1.17}Ga_{0.83}O_3$ sample when bias voltage limit increased from 50 V to 200 V. The essential parameters, obtained by fitting the S2 segment of I-V curve in P mode using Schottky equation (at low voltage regime) and P-F equation (at high voltage regime), are shown in Table 1. The barrier height ($\varphi_B$), calculated using intercept of the linear fit on Y axis, decreased on increasing the bias voltage limit. The barrier height also notably decreased at higher voltage regime of the I-V curve where P-F equation is best fitted in comparison to the values at lower voltage regime where Schottky equation is best fitted (within the same bias voltage limit). The slope ($s_{SE} = \frac{q\sqrt{q}}{k_B T \sqrt{4\pi\varepsilon_0\varepsilon_r}}$) values using Schottky equation are found to be smaller (~ 0.007 for 50 V, 0.005 for 100 V and 0.004 for 200 V limits) for $Cr_{1.45}Ga_{0.55}O_3$ sample in comparison to the values (~ 0.010 for 50 V, 0.006 for 100 V and 0.005 for 200 V limits) for $Cr_{1.17}Ga_{0.83}O_3$ sample. In contrast, slope values using fit of P-F equation ($s_{PF} = \frac{q\sqrt{q}}{rk_B T \sqrt{\pi\varepsilon_0\varepsilon_r}}$) for $Cr_{1.45}Ga_{0.55}O_3$ sample (~ 0.089, 0.028 and 0.009 for limits 50 V, 100 V and 200 V, respectively) are larger in comparison to the values for $Cr_{1.17}Ga_{0.83}O_3$ sample (~ 0.055, 0.020 and 0.007 for limits 50 V, 100 V and 200 V, respectively). The slope values are decreased on increasing the bias voltage limits. It could be due to increase of local temperature from 300 K as used for calculation, especially at higher voltage. The parameter '$r$', using the ratio of $\frac{s_{SE}}{s_{PF}}$, is found the less than 1 for bias voltage limits 50 V and 100 V, and approaching to 1 or greater for bias voltage limit 200 V. The dynamical dielectric constant ($\varepsilon_r$) values are well below of 1, which may not be physically correct for highly resistive metal oxide. In the calculation of $\varepsilon_r (= \frac{q}{4\pi\varepsilon_0}(\frac{q}{k_B T})^2 \times \frac{1}{s^2})$, we have used q as single electronic charge. It may be over simplified considering the fact that electric field induced migration and diffusion of ions play a significant role on charge conduction mechanism, where the net charge of the moving entity is multiple of electronic charge. The Schottky equation and P-F equation were not applicable for NDR regime and above in the S1 segment of I-V curve. Although many metal oxides with M/MO/M structure have shown the coexistence of resistive switching and NDR effect [9, 11, 31], but the exact mechanism is not clear. Jia et al. [20] observed the NDR in negative voltage side of I-V loop. Yang et al. [19] observed resistive switching and NDR in Au/BTO/FTO system, where the behaviour was explained in terms of trapping and de-trapping electrons at the interfaces. The coexistence of RS and NDR in α-Fe$_2$O$_3$ nanorods was attributed to defects states (oxygen vacancies and interstitial Fe ions) [11]. In our samples, the current (I) in the S1 segment (LRS) of P mode after reaching a peak value (I$_P$) at bias voltage V$_P$ started to decrease down to a valley, which is either levelled off or creeping up at higher voltage (V>V$_P$) depending on the limit of bias voltage. The increasing current above valley point attributed to the fact that kinetic energy of the charge carriers may be enough for charge carriers to overcome the barrier height. On the other hand, kinetic energy in the NDR region is not sufficient to overcome the energy barrier at the M/MO interface. The I-V curve of S1 segment in NDR regime and above obeyed $J = AE^2 e^{(\frac{B}{E})}$ equation with positive slope in the plots ln(J/AE$^2$) vs 1/E (insets) for both Cr$_{1.45}$Ga$_{0.55}$O$_3$ (Fig. 5(a-c)) and Cr$_{1.17}$Ga$_{0.83}$O$_3$ (Fig. 5(d-f)) samples. The fitting of I-V curve using Fowler-Nordheim type equation with positive slope was noted in a few semiconductor materials [26, 32]. On the other hand, a true Fowler-Nordheim equation ($J = AE^2 e^{(-\frac{B}{E})}$), generally describes a direct electron tunneling from cathode to conduction band of metal oxide, needs negative slope (-B). This means I-V characteristics in the NDR regime and above are controlled by the transport or hopping of charge carriers through bulk material. We understood that I-V curve in the NDR regime (a relatively low field) is dominated by the term $J \propto e^{(\frac{B}{E})}$ and above NDR regime (relatively high field) is dominated by the term $J \propto E^2$. The plateau type behaviour appears due to the competition between these two terms. The barrier height values, calculated from the fit values of B, are relatively small in comparison to that obtained for S2 segment using P-F equation. Also, the barrier height in NDR regime is found relatively small in comparison to the values at higher voltages above NDR. One indication is that at higher voltages within the same bias voltage limit or by increasing the voltage limit, the space charge effect associated with the electric field induced diffusion of ionic charges inside the lattice structure plays an important role on conduction mechanism.

## 3.2. Effect of measurement delay time on I-V characteristics

The non-equilibrium character in the I-V curves was investigated by varying the delay time between two bias voltage points. The measurement was repeated 20 times by sweeping the bias voltage from -200 V to + 200 V and +200 V to - 200 V using Keithley 6517B electrometer. The I-V curves for $Cr_{1.45}Ga_{0.55}O_3$ sample showed NDR effect at positive voltage side while bias volatge switched from negative to positive (Fig. 6 (a-c)) and vice-versa (Fig. 6 (d-f)) for delay time 0 ms to 500 ms. In case of $Cr_{1.17}Ga_{0.83}O_3$ sample (Fig. 7), the current saturated at higher voltages due to compliance limit, when voltage swept from -200 V to + 200 V at zero delay time. The current value decreased on increasing the delay time to 200 ms and 500 ms. This made the sample in high resistance state (inset of Fig. 7(b)) on repeating the measurement, but there was no clear signature of NDR effect while voltage swept from -200 V to + 200 V for delay time up to 200 ms. The NDR effect appeared at positive side of the bias voltage for delay time at 500 ms (Fig.7(c)). The I-V curves recorded at delay time 0 ms to 500 ms showed *NDR* effect at negative voltage region, when bias voltage swept from $+200\,V\ to-200\,V$. The appearance of NDR on switching the polarity of bias voltage at different delay times depends on the direction of applied bias voltage for $Cr_{1.17}Ga_{0.83}O_3$ sample, unlike an independence of direction of bias voltage for $Cr_{1.45}Ga_{0.55}O_3$ sample. The break down regime appreaed for both the samples when measurement started from higher voltage, but break down regime did not appear on switching the polarity of voltage. Fig. 8 shows the variation of NDR peak position ($V_P$) and height ($I_P$) with the number of repetition of the measurement (N). Although the estimated magnitudes of $V_P$ and $I_P$ are a bit scattered, an increasing trend is noted with the measurement repetition in voltage sweeping modes ±200 V to ∓200V for $Cr_{1.45}Ga_{0.55}O_3$ sample (Fig. 8(a-b)). On the other hand, $V_P$ and $I_P$ values are not much varied with measurement repetition in both voltage sweeping modes for $Cr_{1.17}Ga_{0.83}O_3$ sample (Fig. 8(c-d)). The appearance of NDR effect after changing the polarity of bias voltage, and its dependence on measurement repetition and delay time suggest that non-equilibrium I-V characteristics may be related to a rapid accumulation of charge carriers at the defect sites and M/MO junctions at low voltage region. At higher bias voltage, trapping/de-trapped or relaxation of charge carriers at defects sites control the nature of I-V curves.

### 3.3. Time relaxation of current

In order to understand the relaxation effect on I-V characteristics, we have measured current as the function of measurement time up to 1 hr in the presence of constant 20 V. Fig. 9 shows that current under constant voltage decreases with the increase of measurement time and approaching towards a steady value. The time response of current in the samples is well fitted with an exponential decay function with two components.

$$I(t) = I_0 + A_1 e^{-t/\tau_1} + A_2 e^{-t/\tau_2} \tag{4}$$

Here, $I_0$ is the non-relaxing (residual) current value, $A_1$ and $A_2$ are constants, $\tau_1$ and $\tau_2$ are the relaxation time constants related to two separate relaxation mechanisms. We conclude that there are two relaxation mechanisms, a fast-response at initial stage with time constant $\tau_1$ and a slow-response at later stage with time constant $\tau_2$. We have obtained $\tau_1$= 2.11 s (fast relaxation) and $\tau_2$= 29.94 s (slow relaxation) for $Cr_{1.45}Ga_{0.55}O_3$ sample, and corresponding $\tau_1$ = 1.93 s and $\tau_2$= 21.86 s for $Cr_{1.17}Ga_{0.83}O_3$ sample. The fast response is caused by trapping of the charge carriers at defect sites and the slow-response is attributed to a combined effect of trapping and de-trapping of charge carriers from defect sites and recombination of charge carriers while passing through the bulk sample [33-36].

### 4. Discussion and summary of results

There are some models available in literature [1, 37] that explained resistive switching without involving any oxygen atoms or metallic filament formation in transition metal oxides (semiconductor/insulator). In such models, electro migration force and Joule heating controls the switching mechanism. In most of the metal oxide films, the non-volatile bipolar resistance switching has been explained by electrochemical reactions of oxygen ions/vacancies at the junctions, where Schottky barrier is modified. The polycrystalline samples of $Cr_{2-x}Ga_xO_3$ (x = 0.55, 0.83) system (CrGaO) were prepared by chemical reaction and subsequent thermal annealing at 800 $^0$C. The highly Ga doped sample ($Cr_{1.17}Ga_{0.83}O_3$) has shown higher current in comparison to low Ga content sample ($Cr_{1.45}Ga_{0.55}O_3$). It is worthy to mention that I-V loop in the α-$Cr_2O_3$ sample is negligibly small in comparison to Ga doped α-$Cr_2O_3$ sample, as reported within voltage limit ± 30 V [23] and also true for voltage limit ± 200 V (not shown in figure). The results of the I-V curve analysis also suggest the presence of a significant amount of microscopic level defects (oxygen vacancies-positive charge and oxygen ions-negative charge), which played a crucial role in the observed bipolar resistive switching process. In the paramagnetic state of the material, spin contribution to charge transport mechanism can be neglected and the total current density (J) for electrons contribution can be written as $J_n(x) = J_{nE} + J_{nD}(x)$, $J_{nE} = ne\mu_nE$ and $J_{nD}(x) = + eD_n\frac{dn}{dx}$ [12]. In these expressions, n is the electrons density, D is the diffusion coefficient, and μ is the mobility of electrons, E is the electric field, e is the absolute electronic charge. The transport of heavy ions (oxygen vacancies, and oxygen ions) will be controlled by a competition of (1) electric field induced migration, (2) diffusion, and (3) recombination/relaxation of the charge carriers. The charge conduction mechanism in thin pellet shaped samples can be qualitatively explained by using the concept of Schottky barrier and a combined effect of electron injection, trapping/de-trapping and recombination of electrons at vacancy sites and migration of oxygen ions/vacancies, which has been used for different oxide films [8-9, 11, 15, 38].

Fig. 10 schematically shows the energy level diagram of the Pt/CrGaO/Pt electronic structure in at different stages of the I-V curve in S1 and S2 segments of positive bias mode. When CrGaO is sandwiched between top Pt electrode (TE) and bottom Pt electrode (BE), the Fermi levels of Pt and CrGaO try to attain nearly same Fermi energy (FE) level in the absence of bias voltage (Fig. 10(a)). In the presence of oxygen vacancies, the Fermi level of CrGaO is expected to adjust close to energy level of the oxygen vacancies (defects) [8]. The interfaces between Pt and CrGaO form Schottky junctions with symmetric barrier heights ($\varphi_B$) and depletion region. The symmetric nature of barrier height ($\varphi_B$) and depletion region at the junctions are modified under the application of bias voltage. The bias voltage acts upon Fermi level of the Pt electrode to force the electrons out of equilibrium and triggers the flow of electrons from electrode at higher energy (cathode) to the electrode at lower energy (anode) through the semiconductor. In the positive bias mode of S1 segment, the positive electrode/anode (TE) attracts negative charged carriers (electrons in the conduction band and oxygen ions inside the lattice structure) and negative electrode (BE) attracts positive charged carriers (holes in the valence band and oxygen vacancies inside the lattice structure). The net current flow depends on the injection (tunnelling) of charge carriers (electrons) through bottom Pt/GaCrO junction, flow of charge carriers through the bulk material and out flow of charge carriers through top GaCrO/Pt junction. When the applied voltage increases, number of electrons (n) in the conduction band increased due to in flow of injected electrons. At low bias voltage regime of S1 segment in P mode, the number of thermally activated electrons from valence band to conduction band dominates over the number of injected electrons and the depletion region at the junctions is narrow. The current increases due to transport of the thermally activated electrons ($n_0$) in the conduction band ($J_{nE}$), which are nearly linear to electric field at low bias voltage. As the bias voltage increases, the injected electrons start to dominate over the thermally activated electrons and the current shows a rapid increase before reaching to a peak ($I_p$) at bias voltage $V_p$ (Fig. 10(b)). The injected electrons while travelling through the materials; a part of the injected electrons is trapped at oxygen vacancy ($V_o^+$) or Fe ion ($Fe^{3+}$) sites of the lattice structure and rest of the injected electrons moves forward.

The injected electrons after trapping at $V_o^+$ ($V_o^+ + e^-$) or at $Fe^{3+}$ ($Fe^{3+} + e^-$) sites form small polarons [39-40] and polarons can be localized or hop from one trapping centre to next under electric field. In the *NDR* regime, a fraction of the injected electrons are getting trapped at the $V_o^+$ sites and the rest are transported (small polaron hopping) via Fe sites ($Fe^{3+} + e^- \rightarrow [Fe^{2+}] \rightarrow Fe^{3+} + e^-$) to TE. At the same time, other internal process also comes into play. The migration of oxygen vacancies toward BE increases the space charge barrier width at the TE junction that reduces the out flow of electrons (Fig. 10(c)). The total current is counter balanced by the diffusion components ($J_{nD}(x)$) of oxygen vacancies and oxygen ions due to concentration gradient ($\frac{dn}{dx}$) in the material. A significant compensation of $J_{nE}$ by $J_{nD}(x)$ leads to NDR effect, i.e., decrease of current with the increase of bias voltage. When diffusion of the oxygen ions (move towards anode) becomes comparable to the migration of oxygen vacancies, the I-V curve attains the valley point. On further increase of the bias voltage (above NDR regime), diffusion of oxygen ions dominates over the migration of oxygen vacancies. This results in a reduction of space charge region at the TE junction. The electrons are also de-trapped from oxygen vacancy sites and move toward TE (Fig. 10(d)). This increases the out flow of electrons from material to TE and increase of net current for bias voltage above NDR regime. Now, at the beginning of the decrease of positive bias voltage (S2 segment) at higher voltage side, the population of oxygen ions near to TE and oxygen ion vacancies near to BE is more. The injected electrons will be trapped and all the electrons will not be able to de-trap while bias voltage reduces (Fig. 10(e)). The charge trapping/de-trapping process at high voltages of S2 segment is supported by the HRS and P-F mechanism. The concentration gradient induced diffusion of ions (oxygen ions and vacancies) and their relaxation/recombination inside the bulk material increases the space charge region and Schottky barrier height on lowering the bias voltage. This leads to the decrease of tunnelling of the electrons through top junction and net current flow in the circuit (Fig. 10(f)). On returning back to the zero bias voltage in S2 segment (positive side) (Fig. 10(g)), there will be a minor current flow in the circuit due to diffusion components ($J_{nD}(x)$) of electrons under the potential difference arising from the space charge regions at the TE and BE junctions. This creates a resistance gap at zero bias voltage between S2 (HRS) and S1(LRS) of I-V curve.

On reversing the polarity of the bias voltage to negative side (S3 segment of P mode), the TE becomes cathode and BE becomes anode. When bias voltage increases in the negative direction, stored electrons at TE junction are injected to conduction band and start to flow out of the BE junction, giving rise to LRS in the material. This creates a resistance gap between S2 segment (HRS) and S3 (LRS) segments at zero bias. On increasing the negative bias voltage, the charges will response in reverse direction and it results in the negative current. The details of the I-V loop mechanism in the negative bias mode (S3 and S4 segments) are not described schematically. However, response of the I-V loop in the negative bias voltage is nearly symmetric with respect to that in positive bias for $Cr_{1.45}Ga_{0.55}O_3$ sample in comparison to $Cr_{1.17}Ga_{0.83}O_3$ sample. It depends on internal structure and level of disorder of the samples, whose response to electric field depends on the limit of applied voltage. Under negative bias, the $J_{nE}$ component initially dominates and results in the increase of current till achieving the peak (LRS). On further increase of negative voltage, charges are trapping in the defect sites and it produces NDR effect. At higher negative voltage, the diffusion process of negative ions and oxygen vacancies and de-trapping of electrons will determine the shape of I-V curve. On reversing the bias voltage path in negative side (S4 segment) after reaching the limiting value, initially a large amount of charge was stored at the interfaces of electrodes and material. This starts to decrease with voltage till bias voltage in S4 segment reduces to zero and the material will be in HRS. The S4 path will not follow path S3 due to trapping or recombination of the charge carries at the defective sites, leading to a loop. On reversing the polarity of bias voltage again to positive side (TE anode and BE cathode), some electrons and oxygen ions are already present near to cathode and oxygen vacancies will be present near to anode. This additional (stored) charge will rapidly bring the material in LRS with increasing current at lower voltage limit of the S1 segment of $2^{nd}$ cycle of I-V loop. However, relaxation effect of charge carriers at trapping centres is not completely died down upon reversing the polarity of bias voltage after completing the first cycle. This reduces the net current (increase of resistance) in the S1 segment of $2^{nd}$ cycle and so on. Hence, I-V paths (S1→S2→S3→S4) in $2^{nd}$ cycle differ from the paths in $1^{st}$ cycle and on subsequent repetition.

## 5. Conclusions

This experimental work has recorded some important I-V characteristics in Ga-doped $Cr_2O_3$ oxide, e.g., abnormal resistive switching and NDR. The charge conduction mechanism has been understood by analyzing the S1 and S2 segments of the I-V curve with a schematic energy level diagram. The I-V curves do not obey a typical SCLC mechanism. The I-V curve in NDR regime and above obeyed Fowler-Nordheim type equation with positive exponents. The Schottky mechanism controls the charge transport at lower bias voltages, where as the P-F mechanism (drift/diffusion of ions) controls the I-V curves at higher bias voltage of the S2 segment. A gap was observed in the R-V curves near to zero bias voltage between S1 and S2 or S3 and S4 segments. The existence of resistance gap confirms a non-volatile resistive state, which is associated to the jump of resistance state LRS→HRS or HRS→LRS due to reversing the bias voltage path on same side or change of polarity. The relaxation in resistive switching and NDR effect with repetition and time dependent measurement confirmed the non-equilibrium character. The time dependence of current measurement showed two relaxation processes, a fast relaxation at initial state of non-equilibrium process is involved with trapping of charge carriers and a slow relaxation at the later stage of non-equilibrium process is involved with de-trapping and oxygen ion migration. The non-equilibrium I-V characteristics and nature of I-V loop depend on the content of Ga in the samples.

**Acknowledgment:**

RNB acknowledges research grants from Department of Science and Technology, Ministry of Science and Technology (No. SR/S2/CMP-0025/2011) for carrying out this work.

**References:**

[1] R. Waser, Microelectronic Engineering **86**, 1925 (2009).

[2] M. Rozenberg, M.J. Sanchez, R. Weht, C. Acha, F. G.-Marlasca, and P. Levy, Phy. Rev. B **81**, 115101 (2010).

[3] J. J. Yang, I. H. Inoue, T. Mikolajick, and C. S. Hwang, MRS Bull. **37**, 131 (2012).

[4] J.D. Seok, T. Reji, R.S. Katiyar, J. F. Scott, H. Kohlstedt, A. Petraru, and H.C. Seong, Rep. Prog. Phys. **75**, 076502 (2012).

[5] A. Sawa, materialstoday **11**, 28 (2008).

[6] W. Hu, L. Zou, R. Chen, W. Xie, X. Chen, N. Qin, S. Li, G. Yang, and D. Bao, Appl. Phys. Lett. **104**, 143502 (2014).

[7] S.-C. Chen, T.-C. Chang, S.-Y. Chen, C.-W. Chen, S.-C. Chen, S. M. Sze, M.-J. Tsai, M.-J. Kao, and F.-S. Yeh, Solid-State Electronics **62**, 40 (2011).

[8] D.Y. Guo, Z.P. Wu, L.J. Zhang, T. Yang, Q.R. Hu, M. Lei, P. G. Li, L. H. Li, and W. H. Tang, Appl. Phys. Lett. **107**, 032104 (2015).

[9] C. Kura, Y. Aoki, E. Tsuji, H. Habazaki, and M. Martin, RSC Adv. **6**, 8964 (2016).

[10] C.-Y. Lin, C.-Y. Wu, C.-Y. Wu, C. Hu, and T.-Y. Tseng, J. Electrochem. Soc. **154**, G189 (2007).

[11] Y. Cai, Q. Yuan, Y. Ye, J. Liu, and C. Liang, *Phys. Chem. Chem. Phys.* **18**, 17440 (2016).

[12] R.N. Bhowmik and G. Vijayasri, AIP Advances **5**, 067126 (2015).

[13] X. Zou, H.G. Ong, L.You, W. Chen, H. Ding, H. Funakubo, L. Chen, and J. Wang, AIP Adv. **2**, 032166 (2012).

[14] D. Choi, and C. S. Kim, Appl. Phys. Lett. **104**, 193507 (2014).

[15] M. H. Lee, K. M. Kim, G. H. Kim, J. Y. Seok, S. J. Song, J. H. Yoon, and C. S. Hwang, Appl. Phys. Lett. **96**, 152909 (2010).

[16] B.-S. Lee, B.-Y. Kim, J.-H. Lee, J. H. Yoo, K. Hong, and S. Nahm, Current App. Physics **14**, 1825 (2014).

[17] H. Sun, Q. Liu, S. Long, H. Lv, W. Banerjee, and M. Liu, J. Appl. Phys. **116**, 154509 (2014).

[18] J. Doo Seok, S. Herbert, and W. Rainer, Nanotechnology **20**, 375201 (2009).

[19] G. Yang, C. H. Jia, Y. H. Chen, X. Chen, and W. F. Zhang, J. Appl. Phys. **115**, 204515 (2014).

[20] C.H. Jia, X. W. Sun, G. Q. Li, Y. H. Chen, and W. F. Zhang, Appl. Phys. Lett. **104**, 043501 (2014).

[21] W.-J. Yoon, S.-Y. Chung, P. R. Berger, and S. M. Asar, Appl. Phys. Lett. **87**, 203506 (2005).

[22] K. Zhang, H. Liang, R. Shen, D. Wang, P. Tao, Y. Liu, X. Xia, Y. Luo, and G. Du, Appl. Phys. Lett. **104**, 053507 (2014).

[23] R. N. Bhowmik, K. V. Siva, R. Ranganathan, and C. Mazumdar, J. Magn. Magn. Mater. **432**, 56 (2017).

[24] F.-C. Chiu, Advances in Materials Science and Engineering **2014**, 18 (2014).

[25] E. W. Lim and R. Ismail, Electronics **4**, 586 (2015).

[26] M. Müller, G- X. Miao, and J-S. Moodera, Europhys. Lett. **88**, 47006 (2009).

[27] V. Mikhaelashvili, Y. Betzer, I. Prudnikov, M. Orenstein, D. Ritter, and G. Eisenstein, J. Appl. Phys. **84**, 6747 (1998).

[28] Y. Du, H. Pan, S. Wang, T. Wu, Y. P. Feng, J. Pan, and A. T. S. Wee, ACS nano **6**, 2517 (2012).

[29] J. Sun, C. H. Jia, G. Q. Li, and W. F. Zhang, Appl. Phys. Lett. **101**, 133506 (2012).

[30] D.S. Shang, Q. Wang, L.D. Chen, R. Dong, X.M. Li, and W. Q. Zhang, Phys. Rev. B **73**, 245427 (2006).

[31] X. Gao, Y. Xia, J. Ji, H. Xu, Y. Su, H. Li, C. Yang, H. Guo, J. Yin, and Z. Liu, Appl. Phys. Lett. **97**, 193501 (2010).

[32] A. J. Webb, M. Szablewski, D. Bloor, D. Atkinson, A. Graham, P. Laughlin, and D. Lussey, Nanotechnology **24**, 165501 (2013).

[33] H.Y. Peng, G.P. Li, J.Y. Ye, Z.P. Wei, Z. Zhang, D.D. Wang, G.Z. Xing, and T. Wu, Appl. Phys. Lett. **96**, 192113 (2010).

[34] P.C. Wang, P.G. Li, Y.S. Zhi, D.Y. Guo, A.Q. Pan, J.M. Zhan, H. Liu, J.Q. Shen, and W.H. Tang, Appl. Phys. Lett. **107**, 262110 (2015).

[35] Y.B. Nian, J. Strozier, N.J. Wu, X. Chen, and A. Ignatiev, Phys. Rev. Lett. **98**, 146403 (2007).

[36] C. Rossel, G. I. Meijer, D. Brémaud, and D. Widmer, J. Appl. Phys. **90**, 2892 (2001).

[37] S Blonkowski, and T Cabout, J. Phys. D: Appl. Phys. **48**, 345101 (2015).

[38] A. Shkabko, M.H. Aguirre, I. Marozau, T. Lippert, and A. Weidenkaff, Appl. Phys. Lett. **95**, 152109 (2009).

[39] N. Iordanova, M. Dupuis, and K. M. Rosso, J. Chem. Phys. **122**, 144305 (2005).

[40] R.N. Bhowmik, and A.G. Lone, J. Alloys Compd. **680**, 31 (2016).

Table 1. Calculated Barrier height ($\varphi_B$) at the electrode (Pt)-material (CrGaO) junction using Fowler-Nordheim type equation in the NRD regime and above in S1 segment, and Schottky equation at low voltage regime and Pool-Frankel equation at high voltage regime of the S2 segment of I-V curve in P mode of the samples.

|  | Schottky equation | | P-F equation | | Fowler-Nordheim type equation | |
|---|---|---|---|---|---|---|
| Sample name (bias voltage limit) | $\varphi_B$ (eV) | $\varepsilon_r$ | $\varphi_B$ (meV) | $r = \frac{s_{SE}}{s_{PF}}$ | NDR regime (low voltage) | > NDR regime (higher voltage) |
| $Cr_{1.45}Ga_{0.55}O_3$ (50 V) | 0.49 | 0.04 | 21.8 | 0.04 | 0.86 meV | 1.06 meV |
| $Cr_{1.45}Ga_{0.55}O_3$ (100 V) | 0.48 | 0.08 | 9.8 | 0.08 | 0.81 meV | 1.35 meV |
| $Cr_{1.45}Ga_{0.55}O_3$ (200 V) | 0.51 | 0.13 | 4.4 | 0.13 | 1.41 meV | 2.02 meV |
| $Cr_{1.17}Ga_{0.83}O_3$ (50 V) | 0.43 | 0.02 | 13.5 | 0.02 | 0.53 meV | 0.74 meV |
| $Cr_{1.17}Ga_{0.83}O_3$ (100 V) | 0.44 | 0.06 | 6.6 | 0.06 | 0.78 meV | 1.16 meV |
| $Cr_{1.17}Ga_{0.83}O_3$ (200 V) | 0.43 | 0.08 | 3.4 | 0.08 | 1.12 meV | 1.89 meV |

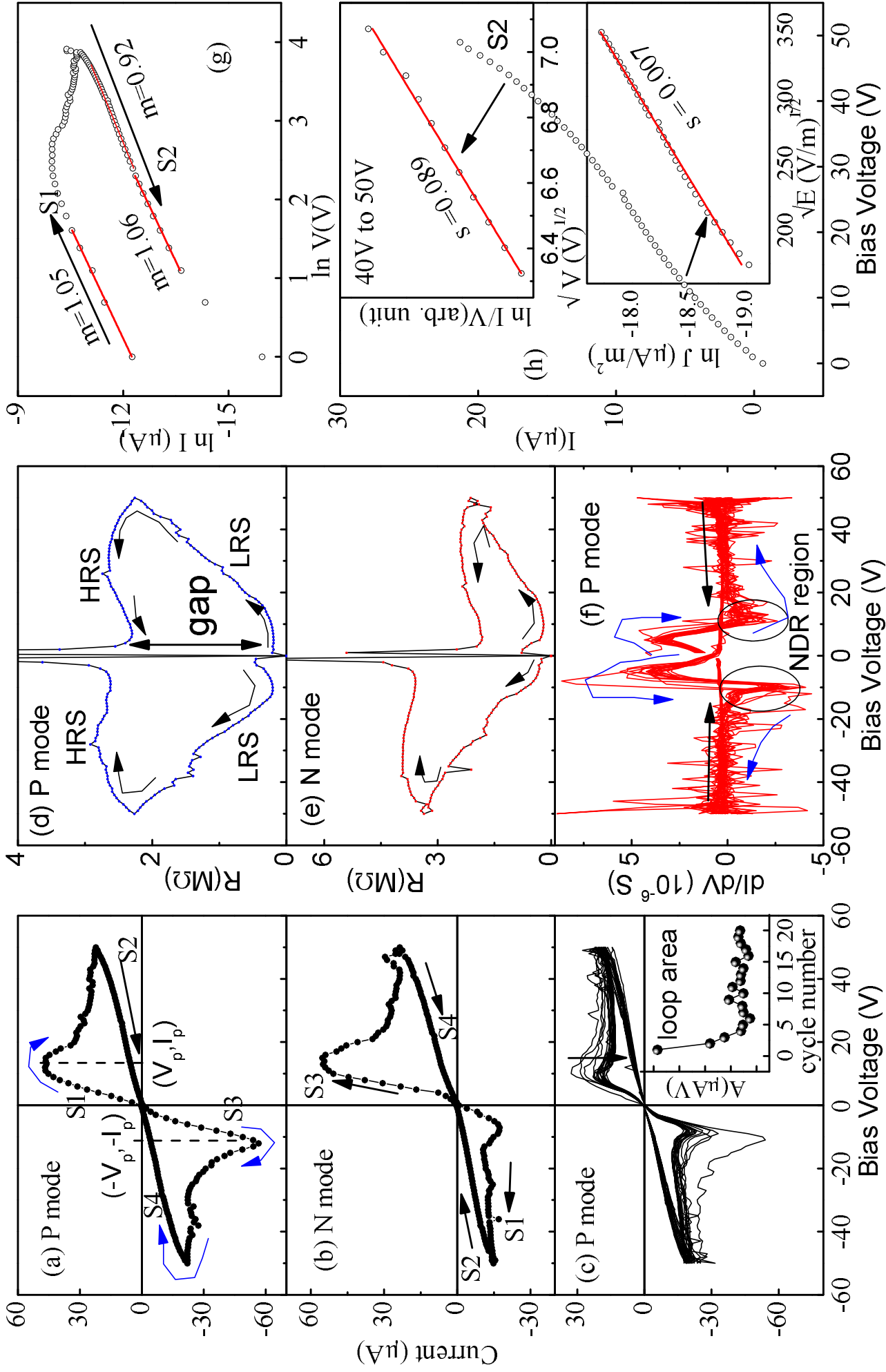

Fig.1 I-V loop for $Cr_{1.45}Ga_{0.55}O_3$ sample within ± 50 V in P mode (a), in N mode (2), voltage cycling effect up to 20 times (c), resistance (R) loop correspondence to I-V data in P mode (d), in N mode (e), differential conductivity data for 20 cycles (f). Power law analysis of S1 and S2 curves in P mode (g), lnJ vs. √E and lnI/V vs. √V for path S2 of P mode (h).

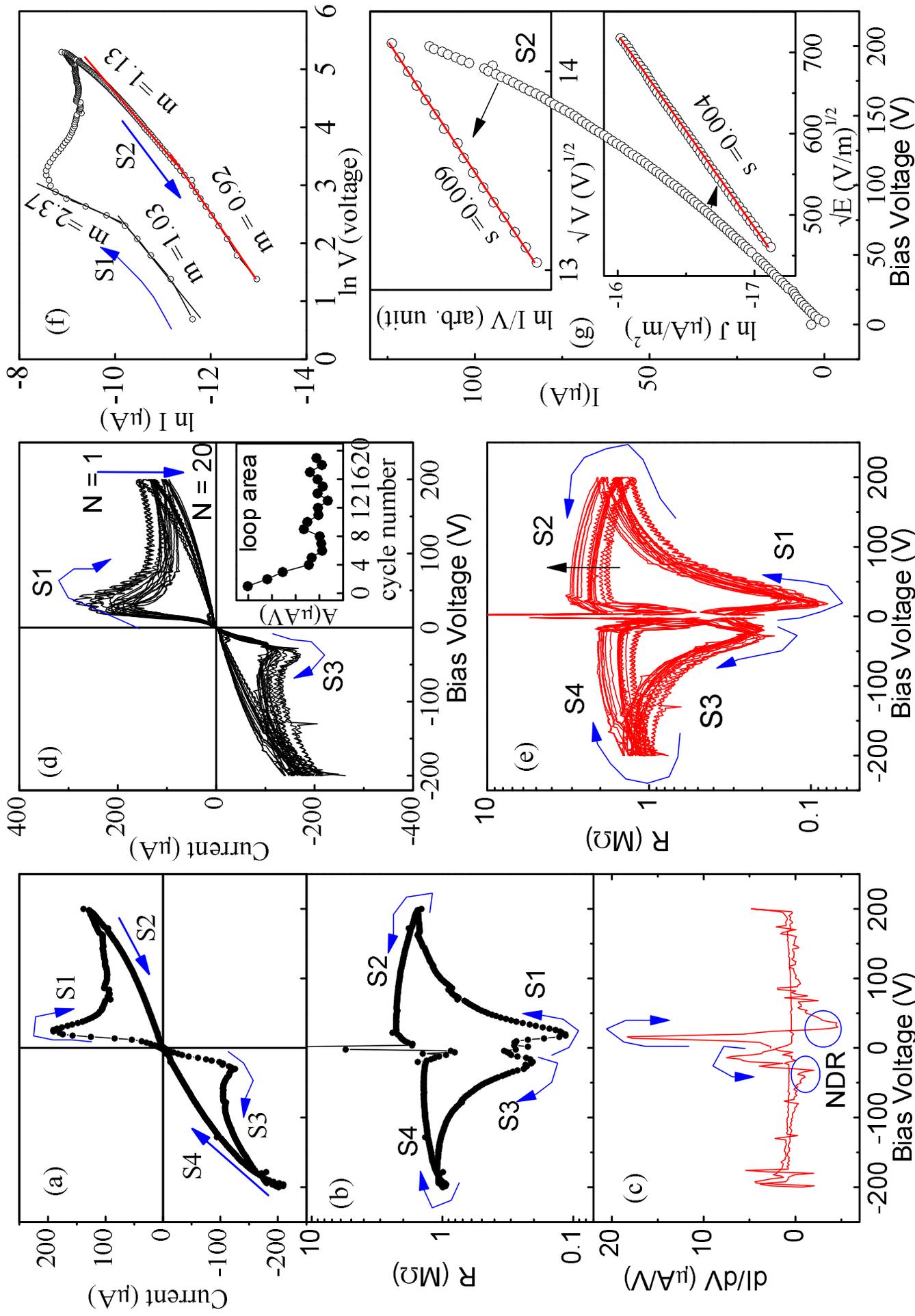

Fig. 2 I-V characteristics of $Cr_{1.45}Ga_{0.55}O_3$ sample at ±200 V in P mode (a) and corresponding R-V loop (b), differential conductance loop (c). Voltage cycling effect on I-V loop (d) and corresponding resistance loops (e). Power law analysis in S1 and S2 curves (f). Analysis of S2 curve using lnJ vs. √E and lnI/V vs. √V plots (insets).

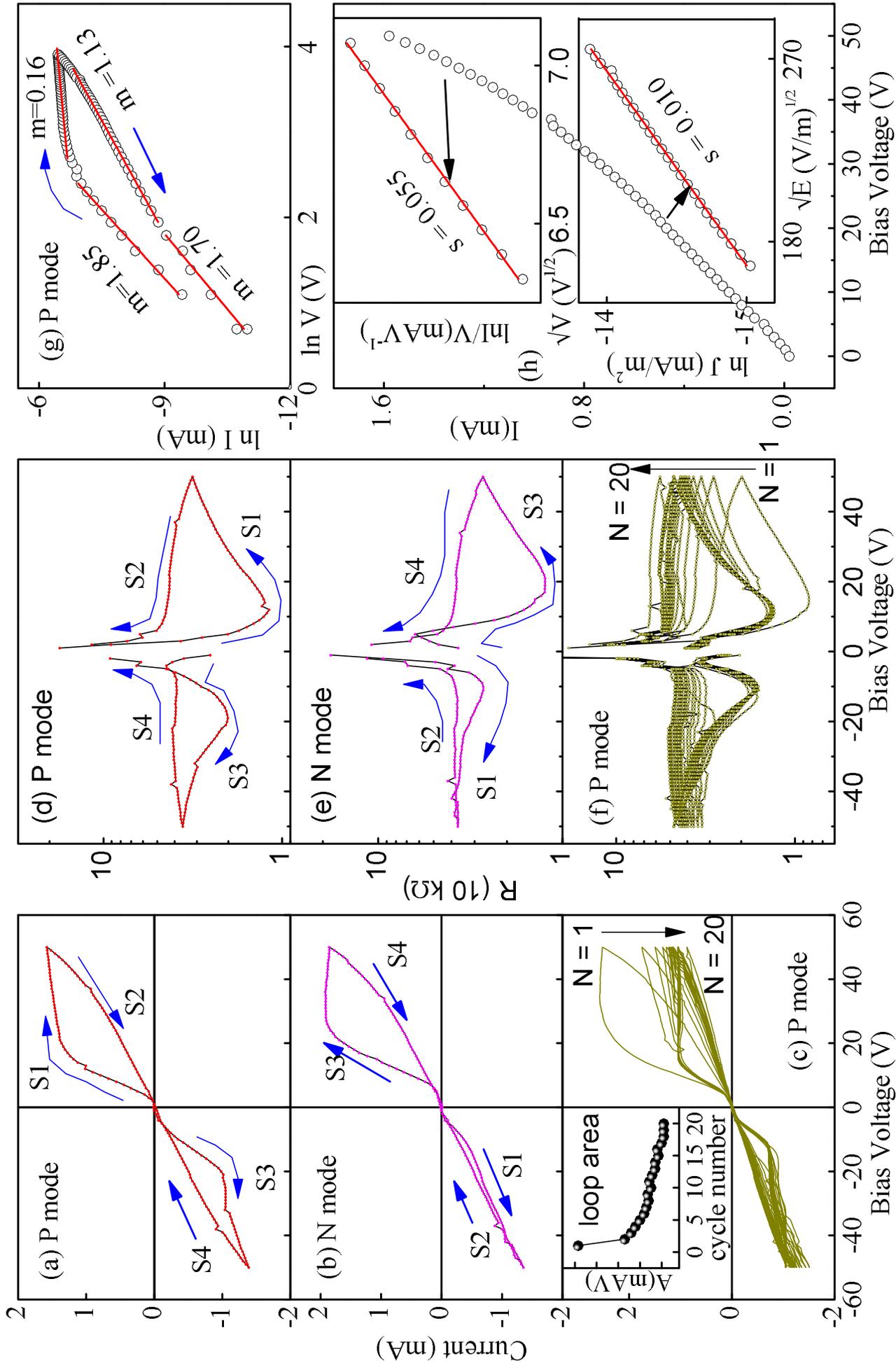

Fig. 3. I-V characteristics of $Cr_{1.17}Ga_{0.83}O_3$ sample at ± 50 V for P mode (a), N mode (b), and voltage cycling effect on the I-V loop in P mode (c). Corresponding R-V loops are shown in (d), (e) and (f), respectively. Power law analysis on S1 and S2 curves of P mode loop (g). Analysis of S2 curve in P mode using ln J vs. √E and lnI/V vs. √V plots (h).

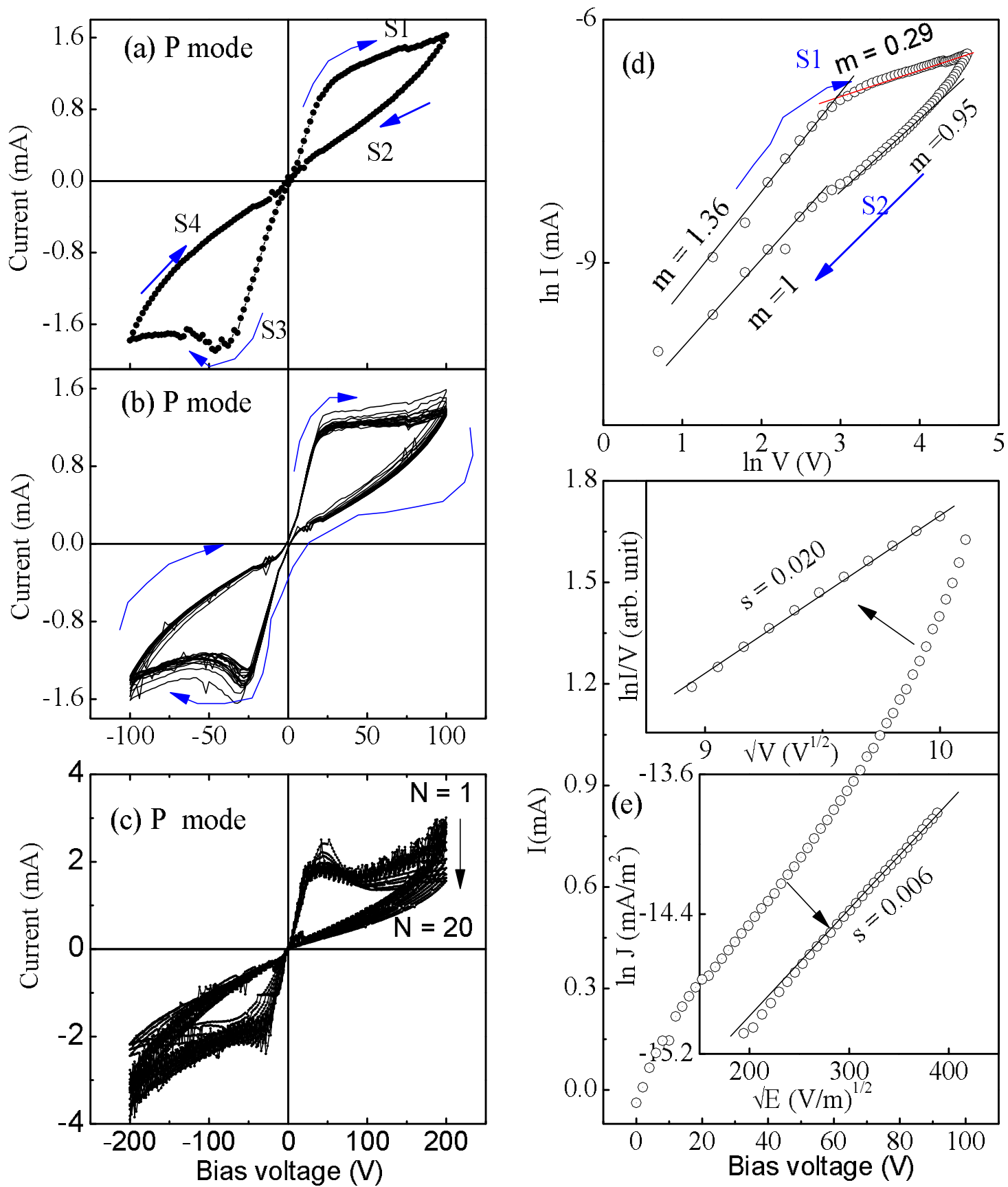

Fig. 4. I-V characteristics of $Cr_{1.17}Ga_{0.83}O_3$ sample at voltage limits ± 100 V (a). Voltage cycling effect on I-V loop within ± 100 V (b) and ± 200 V (c). Analysis for S1 and S2 curves using power law (d) and S2 curve using $\ln J$ vs. $\sqrt{E}$ and $\ln I/V$ vs. $\sqrt{V}$ plots (e) of data from loop in (a).

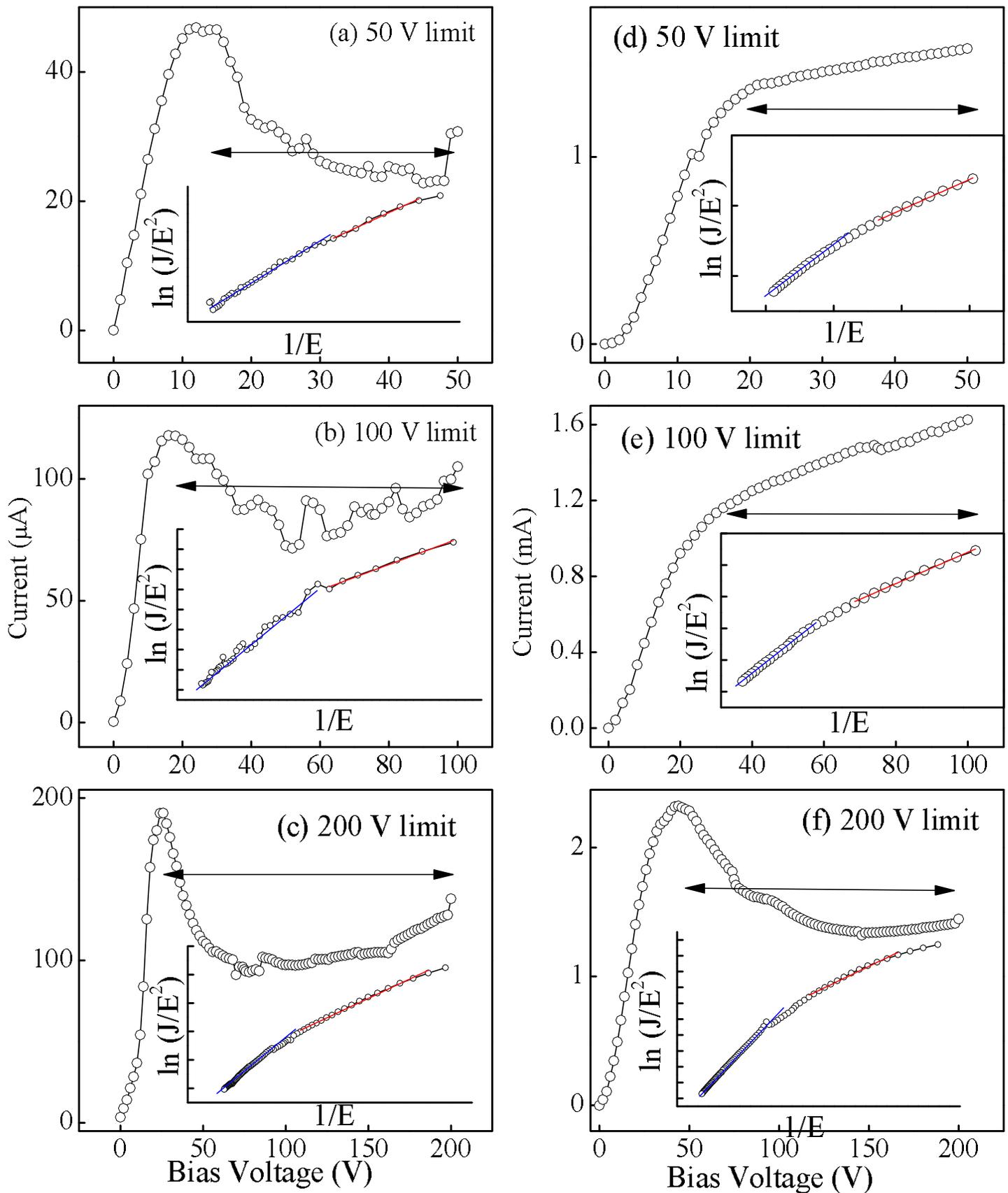

Fig.5 Fit of the I-V curve in NDR regime and above (voltage range is shown by double arrowed line) of $Cr_{1.45}Ga_{0.55}O_3$ (a-c) and $Cr_{1.17}Ga_{0.83}O_3$ (d-f) samples using Fowler-Northeim type equation (inset plots with two slopes). The rage of units of Y and X scales in the insets are not shown for clarity of the figures.

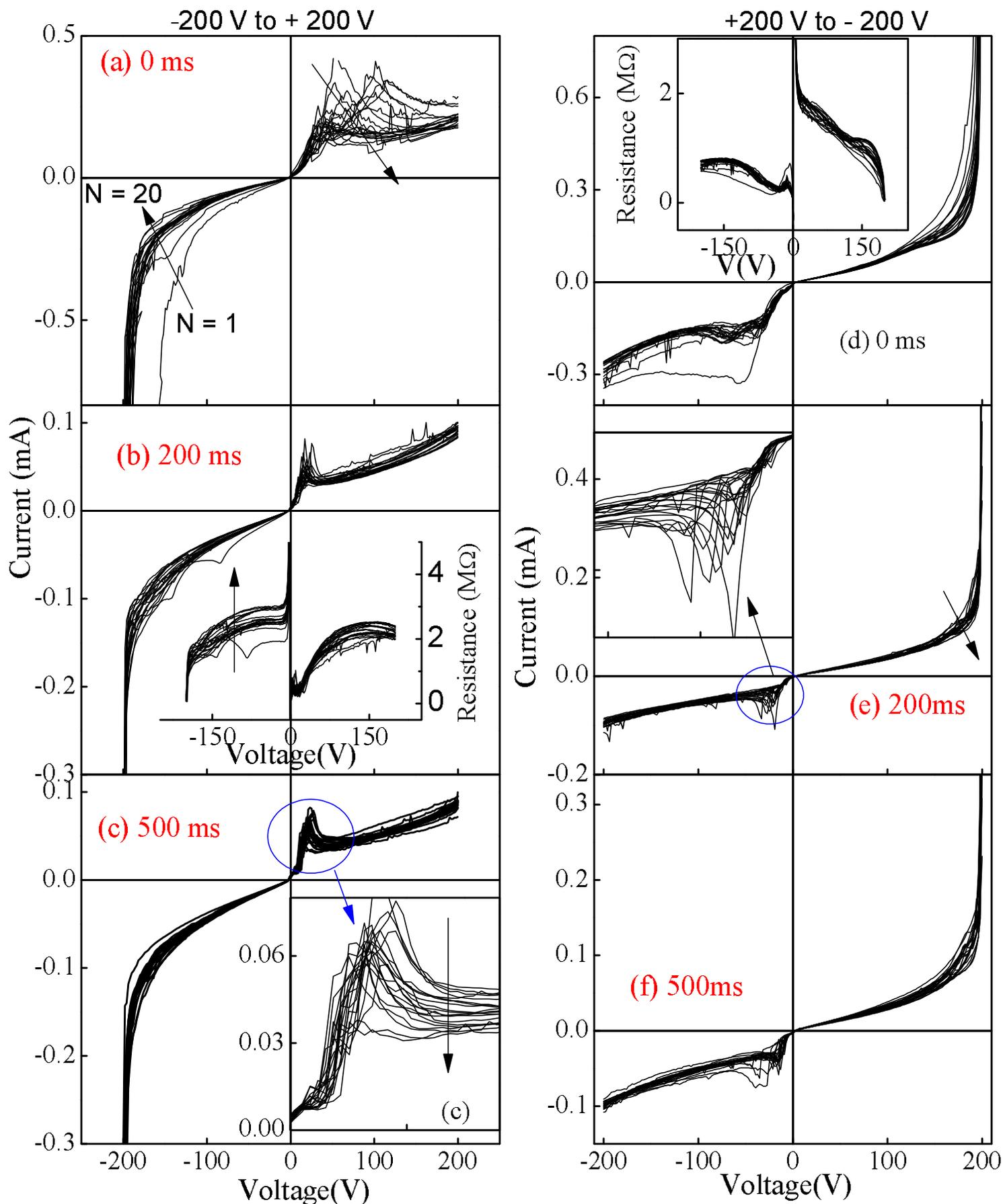

Fig. 6. Repetition of the I-V curves by sweeping the bias voltage from -200V to +200V(a-c) and +200V to -200V(d-f) at different delay time for $Cr_{1.45}Ga_{0.55}O_3$ sample. The associated changes in resistance and NDR peak positions are indicated as inset

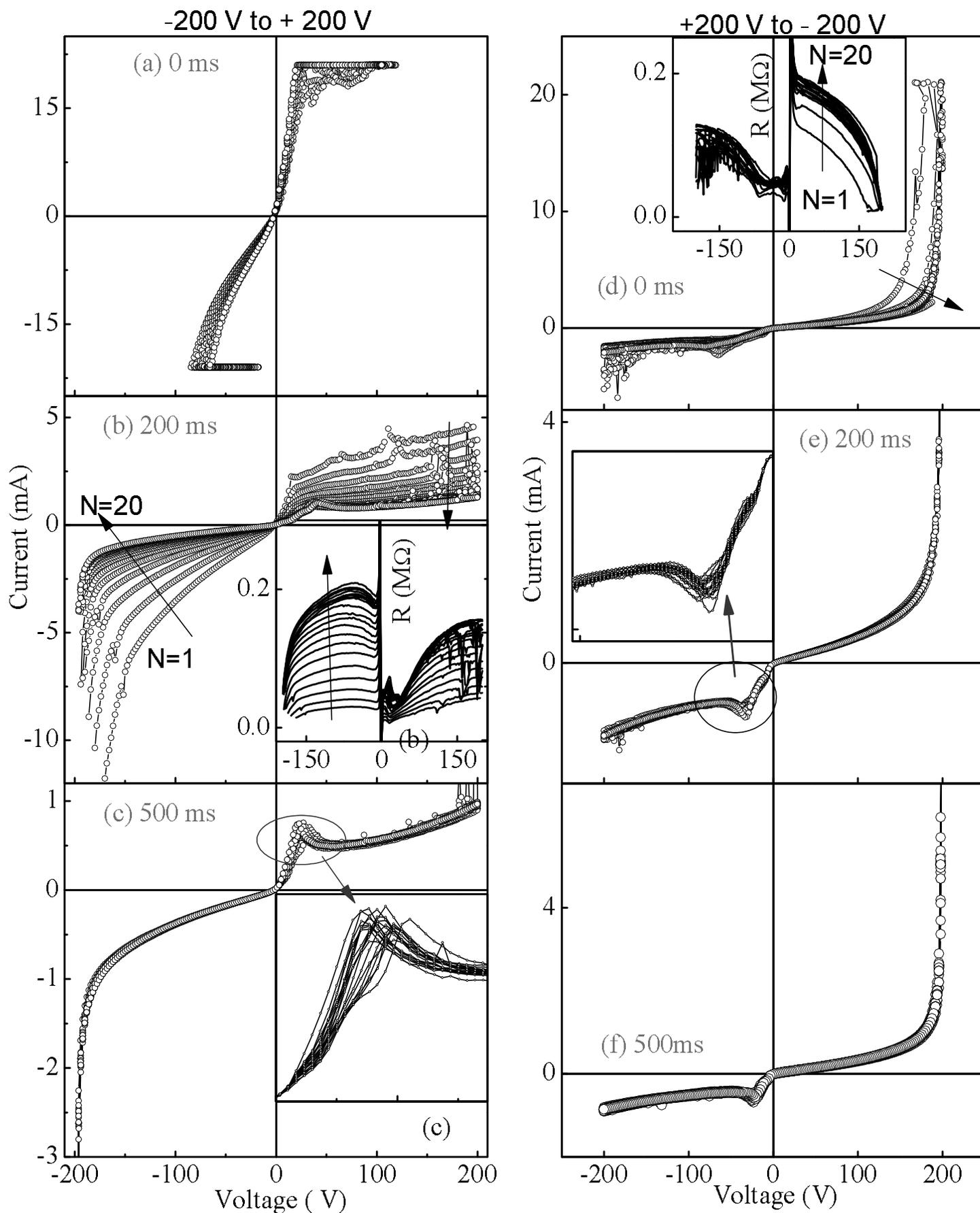

Fig. 7. Repetition of the I-V curves by sweeping the bias voltage from -200V to +200V(a-c) and +200V to -200V(d-f) at different delay time for $Cr_{1.17}Ga_{0.83}O_3$ sample. The associated changes in resistance and NDR peak positions are indicated as inset.

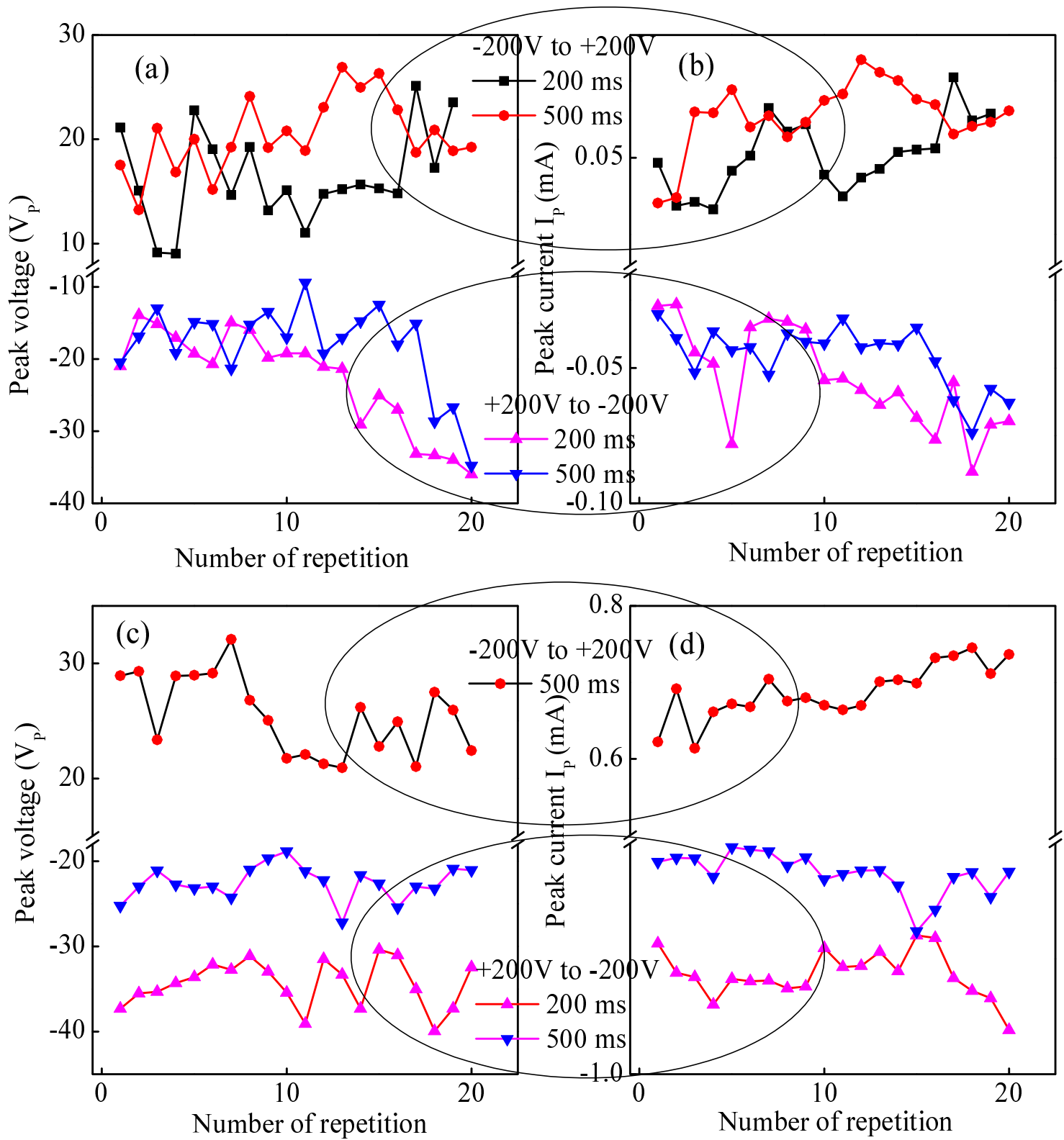

Fig. 8 Variation of peak voltage and peak current values with the number of repetition of measurement for the samples $Cr_{1.45}Ga_{0.55}O_3$ (a-b) and $Cr_{1.17}Ga_{0.83}O_3$ (c-d) for I-V curves recorded at measurement delay times 200 ms and 500 ms.

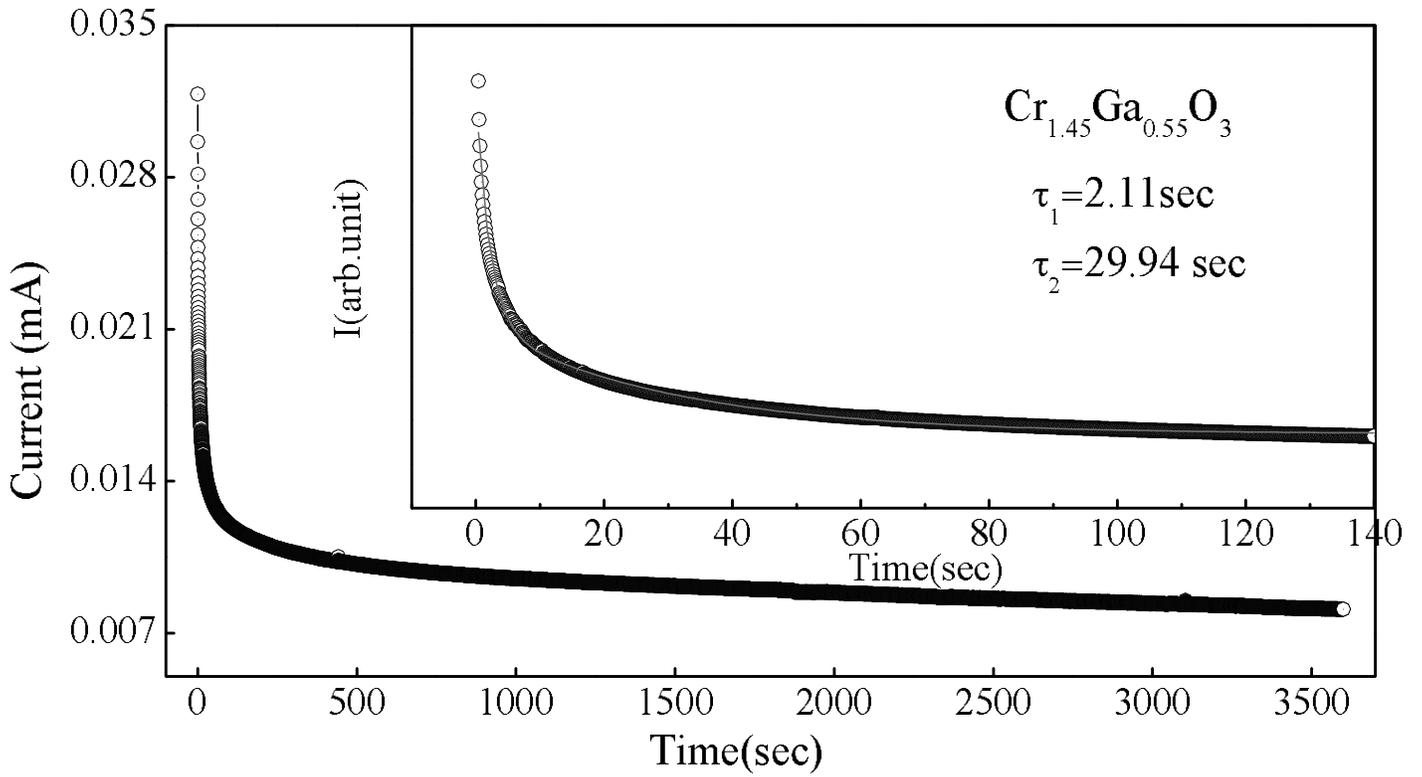
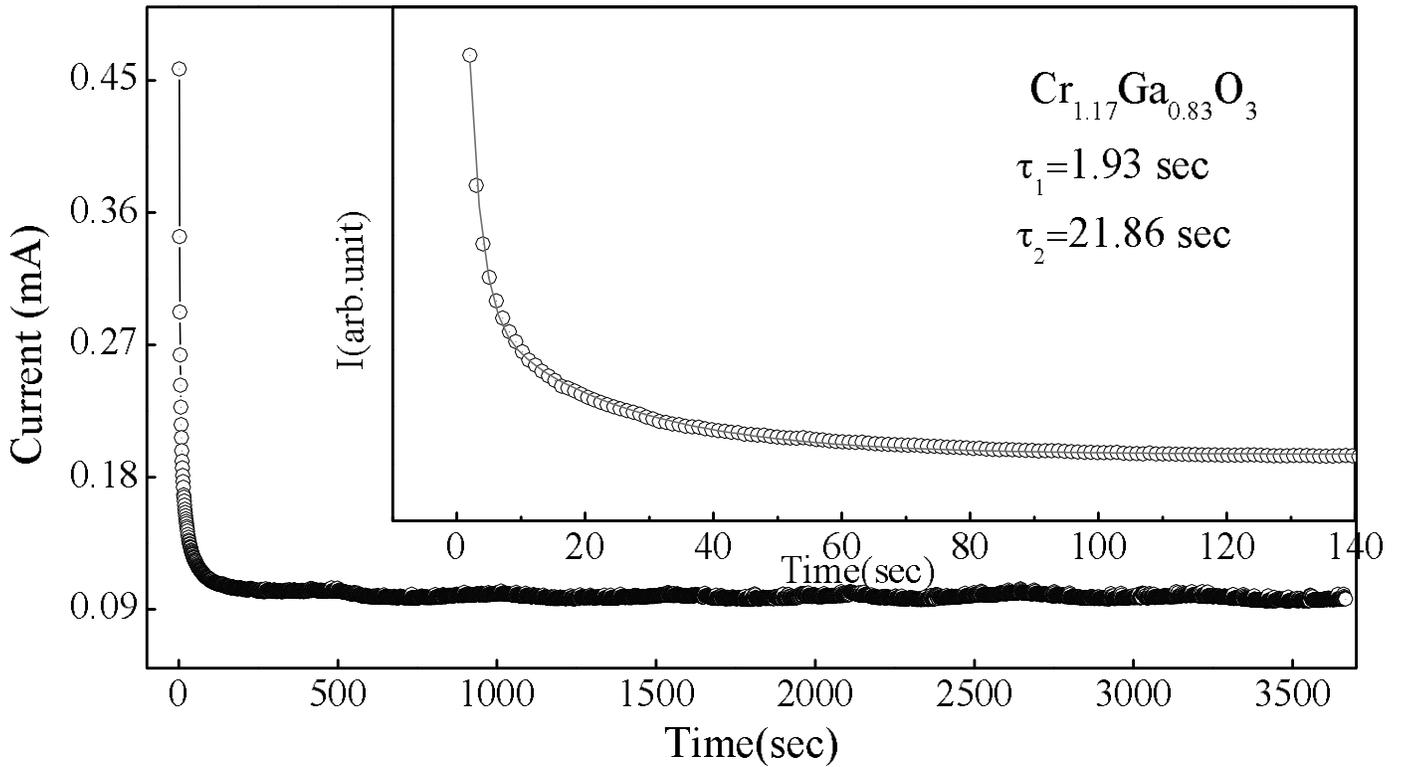

Fig.9. Time dependence of Current measurement at constant voltage for samples $Cr_{1.45}Ga_{0.55}O_3$ (a) and $Cr_{1.17}Ga_{0.83}O_3$ (b), and insert shows the fitting of experimental data.

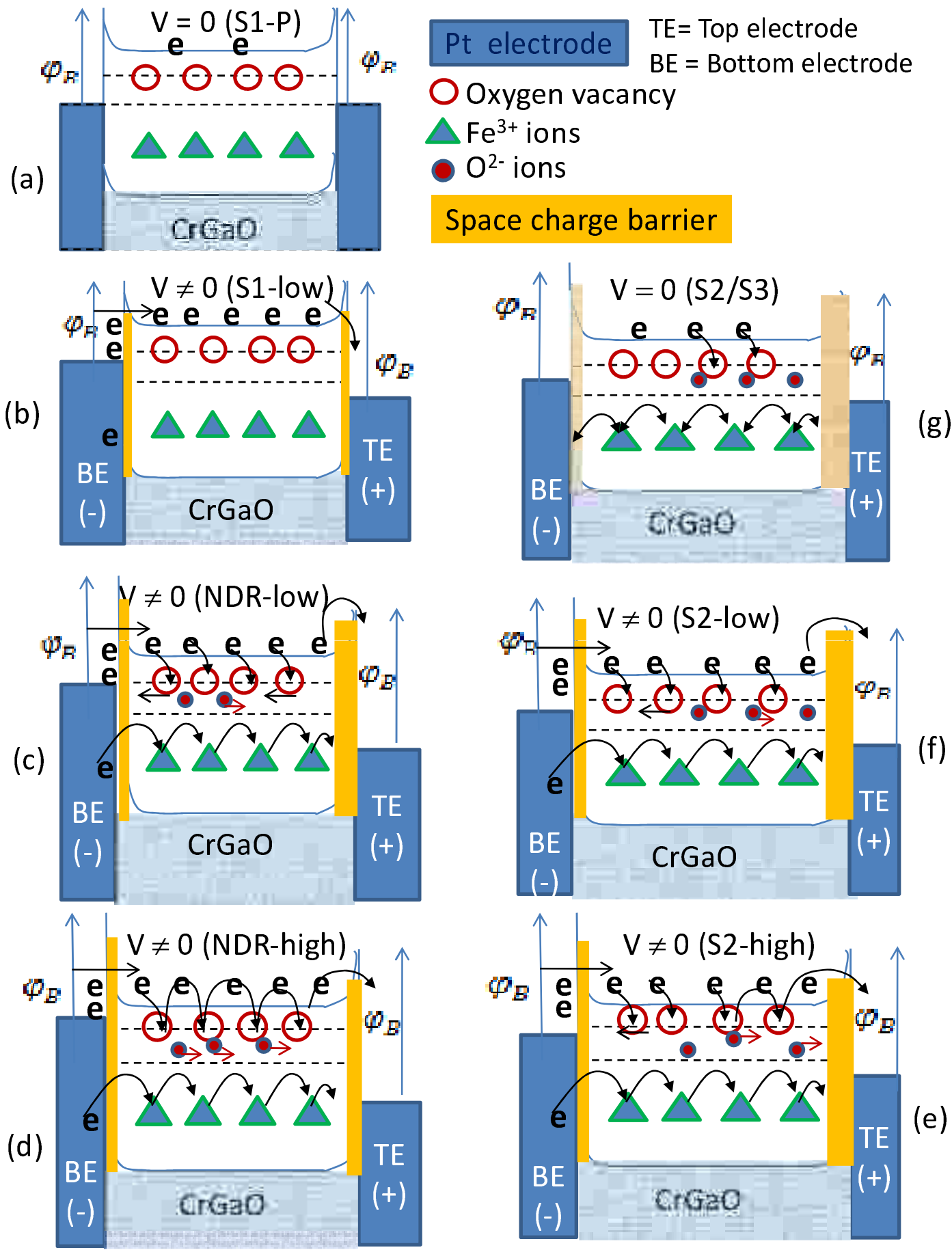

Fig. 10 Schematic energy diagram at different stages of I-V loop at +ve bias mode